# Interplay between Exchange Interaction and Magnetic Shape Anisotropy of ferromagnetic nanoparticles in a non-magnetic matrix for rare-earth-free permanent magnets

Shouvik Sarker[1], Md Mahadi Rajib[1], Radhika Barua[1], Jayasimha Atulasimha[1,2,3,4*]
[1] *Mechanical and Nuclear Engineering, Virginia Commonwealth Univ., Richmond, Virginia USA*
[2] *Electrical and Computer Engineering, Virginia Commonwealth Univ. Richmond, Virginia USA*
[3] *Physics, Virginia Commonwealth Univ. Richmond, Virginia USA*
[4]*Materials Sciences Division. Lawrence Berkeley National Laboratory, Berkeley, CA, USA*
*(affiliation when part of the work was performed, not current affiliation).*
*\*Corresponding author jatulasimha@vcu.edu*

**ABSTRACT**

Developing permanent magnets with fewer critical elements requires understanding hysteresis effects and coercivity through visualizing magnetization reversal. Here, we numerically investigate the effect of the geometry of nanoscale ferromagnetic inclusions in a paramagnetic/non-magnetic matrix to understand the key factors that maximize the magnetic energy product of such nanocomposite systems. Specifically, we have considered a matrix of "3μm×3μm×40nm" dimension, which is a sufficiently large volume, two-dimensional representation considering that the ferromagnetic inclusions thickness is less than 3.33% of the lateral dimensions simulated. Using this approach that is representative of bulk behavior while being computationally tractable for simulation, we systematically studied the effect of the thickness of ferromagnetic strips, the separation between the ferromagnetic strips due to the nonmagnetic matrix material, and the length of these ferromagnetic strips on magnetic coercivity and remanence by simulating the hysteresis loop plots for each geometry. Furthermore, we study the underlying micromagnetic mechanism for magnetic reversal to understand the factors that could help attain the maximum magnetic energy densities for ferromagnetic nanocomposite systems in a paramagnetic/non-magnetic material matrix. In this study, we have used material parameters of an exemplary Alnico alloy system, a rare-earth-free, thermally stable nanocomposite, which could potentially replace high-strength NdFeB magnets in applications that don't require large energy products. This can stimulate further experimental work on the fabrication and large-scale manufacturing of RE-free PMs with such nanocomposite systems.

## INTRODUCTION

Permanent magnets (PMs) are essential in numerous applications, from powering electric motors in home appliances to enabling the operation of nanoscale devices like Hall-effect sensors and data storage [1,2]. Within the energy sector industrial base, and clean energy in particular, PMs are key components of wind turbine generators (especially for offshore turbines) and traction motors in battery and hybrid electric vehicles as they demonstrate an intrinsic large magnetic induction for developing torque ($B$), and a high resistance to demagnetization effects ($H_c$) [3,4]. To date, the most popular PMs currently in the market, NdFeB and SmCo, demonstrate a $(BH)_{max}$ in the range of 30-40 MGOe at room temperature; however, a significant drawback of these magnets is that they contain expensive rare-earth elements, mainly neodymium (Nd), dysprosium (Dy), samarium (Sm), terbium (Tb), and praseodymium (Pr) which are connected to unstable markets, sensitive geopolitical situations, and environmentally risky mining practices [5]. Among the possible alternatives for rare-earth magnets, Alnico alloys (so-called by their majority composition of aluminum, nickel, iron, and cobalt) are attractive as they are comprised entirely of earth-abundant elements that can be recycled easily. Further, they demonstrate high saturation magnetization and high-temperature performance with a small magnetic remanence temperature coefficient (-0.025%/°C) and positive coercivity temperature coefficient (+0.01%/°C) up to 550°C, as well as resistance to corrosion effects [6-8]. The distinguishing feature of Alnico alloys is that they are "fine-particle" alloys, possessing a microstructure consisting of micron or submicron-scale ferromagnetic particles dispersed in a weakly magnetic matrix [9]. Basic Alnico alloys derive their magnetic strength by phase separation in the cast alloy into ferromagnetic FeCo-rich ($\alpha_1$ phase) and a paramagnetic NiAl-rich phase ($\alpha_2$ phase) precipitated from the high-temperature homogeneous composition during spinodal decomposition at elevated temperature [10]. Alnico components fabricated using conventional melt-based processes (casting, melt-spinning) and laser-based additive manufacturing techniques (powder bed fusion additive manufacturing, direct energy deposition processing) are subjected to an essential post-process multi-step heat treatment involving solutionization, annealing (in a magnetic field), and finally tempering, resulting in a microstructure wherein the ferromagnetic $\alpha_1$ precipitates form an aligned array of nanoscaled compass needles in the largely non-magnetic $\alpha_2$ matrix[11-15]. These are the reasons for the choice of material parameters of an exemplary Alnico alloy system for the micromagnetic study in this paper. However, the study applies to any ferromagnetic nanocomposite in a non-magnetic/paramagnetic matrix.

In permanent magnets, the strength is quantified by the maximum energy product (BH)max, a figure of merit which is the product of the remanence Br and the coercivity Hc in the second quadrant of the B-H hysteresis loop (also known as demagnetization curve) shown in Fig 1(a). A good permanent magnet has a magnetization "locked in" which cannot be easily removed by applying a magnetic field. Simply put, the larger the coercive field (Hc) needed to drive the magnetization to zero, the better the magnet; and the permanent magnet needs to produce a large field when no external field is applied to it, i.e., the remnant magnetization (Br) at zero applied field should be as large as possible.

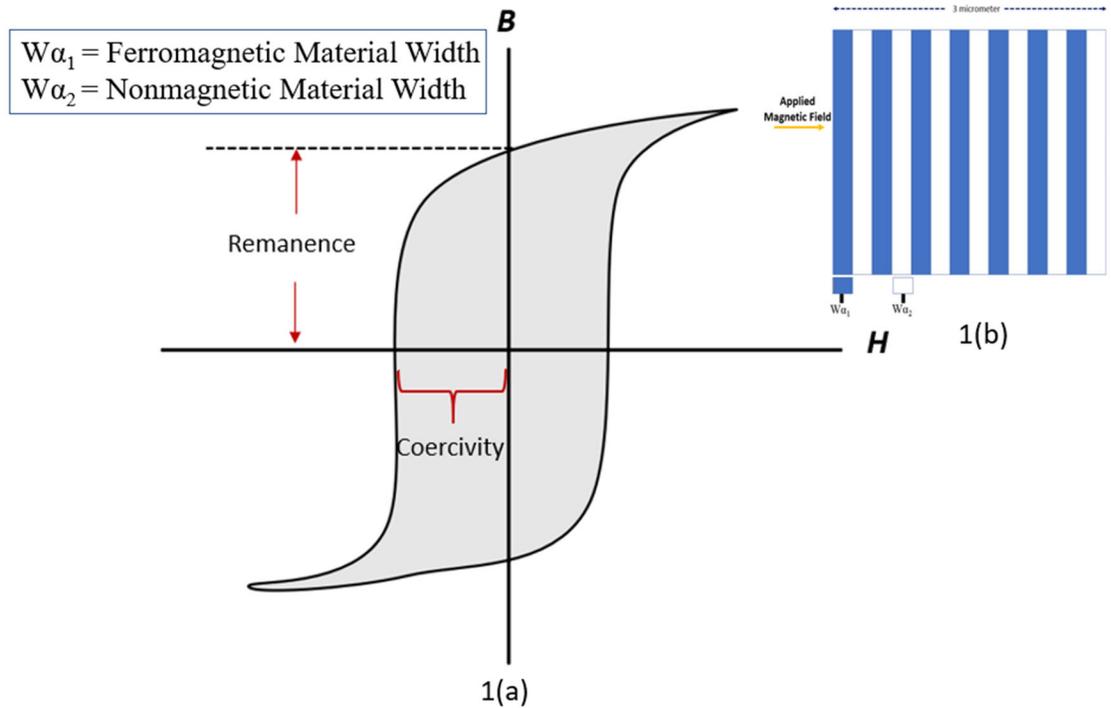

**Fig 1(a).** Magnetic hysteresis loop for a ferromagnet from which remanence, $B_r$, and coercivity, $H_c$ are calculated their product gives the maximum energy product $(BH)_{max}$. **1(b).** Microstructure of the simulated matrix where $W\alpha_1$ represents ferromagnetic material width and $W\alpha_2$ represents nonmagnetic material width.

In this study, we focus on using the micromagnetic simulation software Mumax3 to investigate the influence of geometry and packing fraction of the α1 and α2 phases on the maximum energy product (BH)max of the phase-separated alloy. Following the microstructure of commercially available Alnico 8 alloys [19], the system is modeled as alternating parallel ferromagnetic strips and non-magnetic strips, creating unique "anisotropic" magnets, Fig 2. Here, we have varied the width of the ferromagnetic strip and distance between neighboring strips inside a precisely defined matrix, simulated the change in the behavior of magnetization in the presence of an external magnetic field, and plotted the hysteresis loop. This was used to determine the coercivity Hc, remanence Br, and maximum energy products (BH)max for different geometries to attain the maximum magnetic energy densities for a ferromagnetic nanocomposite system in a paramagnetic/non-magnetic matrix. Further, the micromagnetic mechanism for the magnetization reversal and the role of thermal noise are investigated. Based on first-order simulations (with simple assumptions), this work provides insights that could stimulate further experimental work.

**MICROMAGNETIC SIMULATION SETUP**
We used micromagnetic simulation to study the magnetization dynamics and obtain coercivity ($H_c$) and remanence ($B_r$). We simulated a volume of 3μm×3μm×40nm dimensions and considered a constant cell size of 2.93×2.93×2.5 nm³. We defined ferromagnetic material (marked as "α1") and airgap (marked as

"α2") alternatively in the matrix systematically varied the width of those two layers and simulated the hysteresis loops. In the simulation, we considered exchange stiffness, A=10pJ/m, saturation magnetization, $M_s$=1600kA/m, and damping factor, α=0.1 for the ferromagnetic region and $M_s$=0 for the non-magnetic region (this is a reasonable assumption for a paramagnetic material as well). The ferromagnetic spins were initialized in the +y axis (i.e., along [0,1,0]), and a magnetic field was swept from 500 mT to -500 mT with a step size of 10 mT to obtain the hysteresis loops. The magnetization dynamics of the rectangular-sized matrix are simulated by solving the Landau–Lifshitz–Gilbert (LLG) equation using MuMax3:

$$\frac{d\vec{M}}{dt} = -(\gamma \vec{M} \times \vec{H}_{eff}) - \frac{\alpha \gamma}{M_s}[\vec{M} \times (\vec{M} \times \vec{H}_{eff})] \quad (1)$$

where $\vec{H}_{eff}$ is the effective magnetic field on the ferromagnetic element, which is the derivative of the total energy of that element's magnetization. Accordingly,

$$\vec{H}_{eff} = -\frac{1}{\mu_o \Omega}\frac{dE}{d\vec{M}} \quad (2)$$

where $\mu_o$ is the permeability of the vacuum and E is the total free energy (not energy density), of a particular multiferroic element of volume Ω in the chain. The total free energy of any element in this chain is given by:

$$E = E_{dipole} + E_{shape\ anisotropy} \quad (3)$$

Here, $E_{dipole}$ is the dipole-dipole interaction energy due to the magnetic interactions between nearest neighbor dipoles within the ferromagnetic matrix, influencing the overall magnetic configuration and stability of the material and $E_{shape\ anisotropy}$ represents the energy arising from the geometric shape of the ferromagnetic elements. The rectangular shape induces anisotropy, causing the material to prefer magnetization along certain directions, which affects the magnetization dynamics and hysteresis behavior. By integrating these energy components into our simulations, we could precisely model the impacts of geometric anisotropy and dipole interactions on the magnetization characteristics. This allowed us to examine the implications of the geometry of the ferromagnetic inclusions on the coercivity and remanence and, consequently, maximum energy products of various configurations.

We introduce a thermal field $H_{th}$ into the effective magnetic field $H_{eff}$ to incorporate thermal noise in our micromagnetic simulations. The modified LLG equation, including thermal noise, is given by:

$$\frac{d\vec{M}}{dt} = -[\gamma \vec{M} \times (\vec{H}_{eff} + \vec{H}_{th}) - \frac{\alpha \gamma}{M_s}[\vec{M} \times (\{\vec{M} \times (\vec{H}_{eff} + \vec{H}_{th})\}] \quad (4)$$

The thermal field $H_{th}$ is modeled as a Gaussian white noise with the following properties:

$$Zero\ Mean:\ \vec{H}_{th} = 0 \quad (5)$$

$$Variance:\ \vec{H}_{th,i}(t)\vec{H}_{th,j}(t^*) = \frac{2\alpha k_B T}{\gamma M_s V}\delta_{ij}\delta(t-t^*) \quad (6)$$

where $k_B$ is the Boltzmann constant, $T$ is the absolute temperature, $\gamma$ is the gyromagnetic ratio, $M_s$ is the saturation magnetization, $V$ is the volume of the computational cell, and $\delta$ represents the Dirac delta function.

## RESULTS AND DISCUSSIONS

To obtain the maximum energy products for the rectangular matrix, we simulated the magnetization dynamics by varying the width and length of the ferromagnetic and nonmagnetic materials. We studied the strength of different geometries by **(i)** varying the width of the ferromagnetic region ($W\alpha_1$) while keeping the nonmagnetic region width ($W\alpha_2$) fixed at 40 nm, **(ii)** varying the width of the nonmagnetic region ($W\alpha_2$) while keeping the ferromagnetic width constant at 40 nm (for case (i) & (ii) length of the ferromagnetic and nonmagnetic material is constant at 3 μm), and **(iii)** keeping both the ferromagnetic and nonmagnetic material width fixed at 40 nm while changing the length of the ferromagnetic strip. In all cases, we initiate the simulations from a purely ferromagnetic state (all the spins pointing in the + y direction).

In Fig 2, the blue dotted line represents a hysteresis loop of width 10 nm ferromagnetic strip ($W\alpha_1$). Here, the nonmagnetic material width ($W\alpha_2$) is constant at 40 nm for all loops. As we increase the ferromagnetic strip width ($W\alpha_1$) from 20 nm up to 100 nm with an increment of 20 nm (represented by orange, yellow, violet, and pink solid lines, respectively), a general trend in the hysteresis loop is observed where with the increase of $W\alpha_1$, the loops get thinner representing a decrease in coercivity. As the ferromagnetic material width ($W\alpha_1$) increases, the internal opposition to spin reversal lowers due to larger dipole coupling that favors the spins to be antiparallel, making it easier for the external magnetic field to align or reverse them. However, with increasing $W\alpha_1$ (fig 2), the packing fraction increases, and as a result, the hysteresis loop gets longer, resulting in higher remnant values. In other words, the increase in effective $M_s$ due to a higher packing fraction is more significant than the reduction in remnant magnetization ($M_R$) due to dipole coupling, resulting in a higher $M_R$. Thus, an optimum ferromagnetic inclusion thickness (40 nm) results in $(BH)_{max}$ for a fixed non-magnetic gap of 20 nm.

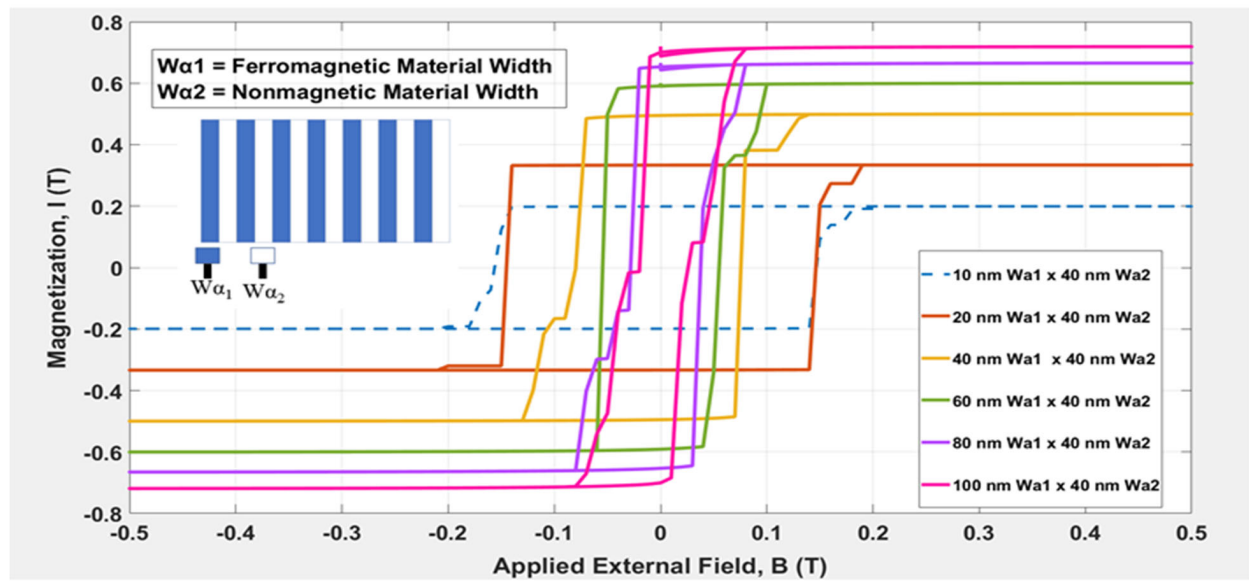

**Fig 2.** Hysteresis loops for a fixed nonmagnetic material width ($W\alpha_2$=40 nm) and varying ferromagnetic strip width ($W\alpha_1$)

After observing the trend of hysteresis loops with varying $W\alpha_1$, we tried to find the changes in the pattern of hysteresis loops with varying $W\alpha_2$. For this study, the ferromagnetic material width ($W\alpha_1$) is fixed at 40 nm (Fig 3.), and we vary the nonmagnetic material width ($W\alpha_2$) from 20 nm up to 100 nm with an increment of 20 nm.

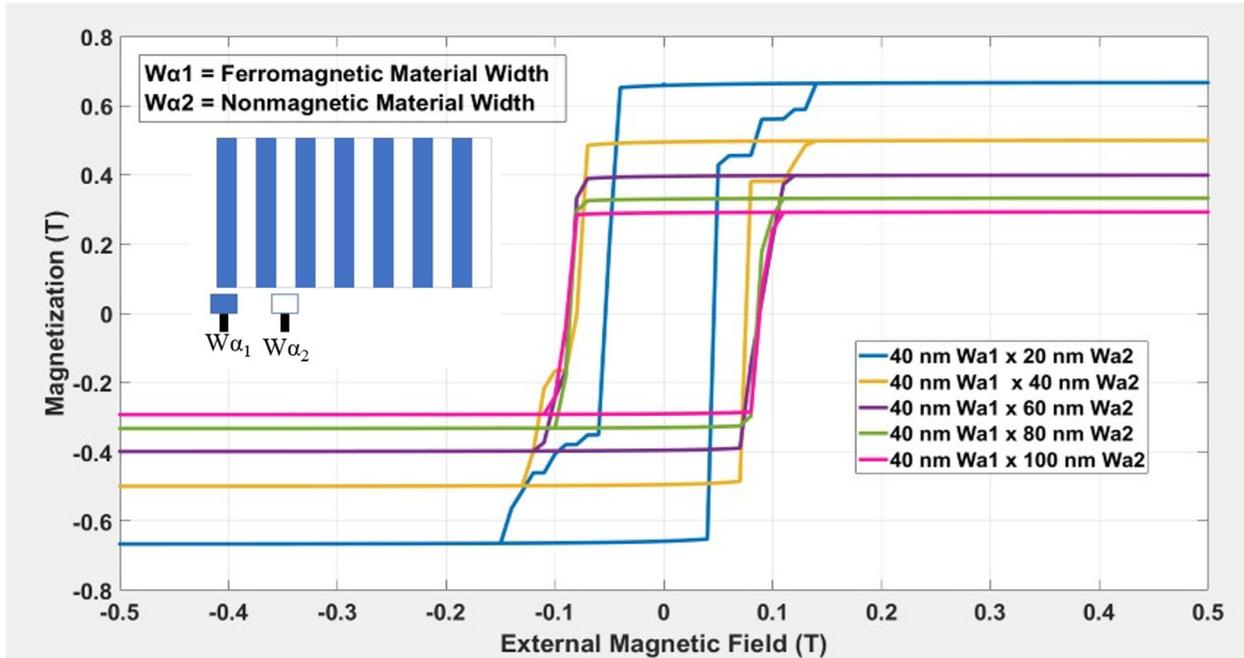

**Fig 3.** Hysteresis loops for a fixed ferromagnetic strip width ($W\alpha_1$=40 nm) and varying nonmagnetic material width ($W\alpha_2$), where $W\alpha1$ = 20,40,60,80,100 nm represented by the blue, yellow, violet, green, and pink solid lines respectively.

Here, the blue solid line represents the minimum nonmagnetic gap of 20 nm, where the hysteresis loop has the highest remanence and lowest coercivity value. When the nonmagnetic material width ($W\alpha_2$) is smaller, it promotes more substantial spin alignment across interfaces, enhancing remanence by preserving more aligned spins. On the other hand, in smaller $W\alpha_2$ regions, the coercivity is lower as the energy barrier for magnetization reversal is low.

From Fig 4, it can be seen that the greatest energy product is achieved when the width of $W\alpha_1$ and $W\alpha_2$ is nearly 1:2. As the ferromagnetic region width increases from 10 to 20 nm, the energy product increases and around the packing fraction of 1:2, the energy product is the maximum. If we compare the maximum energy product of the RE magnets with our simulated results, the $(BH)_{max}$ of RE magnets used practically falls in the range of 25-30 MGOe, whereas, in our case, the maximum achievable $(BH)_{max}$ is around 5 MGOe which is around 20% of RE magnets. Although it is small compared to RE magnets, we can improve the values and try to reach within 50% of the RE magnets' energy products by implementing negative exchange bias and introducing edge roughness (not taking into consideration in our ideal simulation result).

**Table 1:** Values of coercivity $H_{ci}$, remanence $B_r$, and energy product $(BH)_{max}$ for different cases. Here, the ferromagnetic material width is denoted as $W\alpha_1$, and the nonmagnetic material width is denoted as $W\alpha_2$.

| Dimension(nm) ($W\alpha 1 \times W\alpha 2$) | Coercivity, $H_{ci}$ (KA/m) | Remanence, $B_r$(T) | Energy Product $(BH)_{max}$ (KJ/m³) | Energy Product $(BH)_{max}$ (MGOe) |
|---|---|---|---|---|
| 10 × 40 | 115.6 | 0.1988 | 22.98 | 2.888 |
| 20 × 40 | 115.4 | 0.3334 | 38.47 | 4.834 |
| 40 × 40 | 61.6 | 0.4950 | 30.492 | 3.832 |
| 60 × 40 | 43.8 | 0.5910 | 25.89 | 3.253 |
| 80 × 40 | 19.6 | 0.6532 | 12.80 | 1.608 |
| 100 × 40 | 19.1 | 0.7006 | 13.38 | 1.681 |
| 40 × 20 | 37.8 | 0.6585 | 24.89 | 3.128 |
| 40 × 60 | 67.64 | 0.3957 | 26.77 | 3.364 |
| 40 × 80 | 68.50 | 0.3300 | 22.61 | 2.841 |
| 40 × 100 | 69.45 | 0.2904 | 20.17 | 2.535 |

**Table 1** quantitatively represents Coercivity $H_{ci}$, remanence $B_r$, and energy product $(BH)_{max}$ from the hysteresis loops for different cases. Among all the geometries, the geometry of 20nm × 40nm $W\alpha_2$ produces the maximum energy product of 38.47 kJ/m3 or 4.834 MGOe. The table shows that as the ferromagnetic strip width ($W\alpha_1$) increases, the coercivity goes down significantly. Hence, we see a decreasing trend of the maximum energy product even though the $M_R$ is higher when the value of nonmagnetic material width ($W\alpha_2$) is fixed, and ferromagnetic material width ($W\alpha_1$) is varied, the energy product goes down with increasing width of the nonmagnetic material.

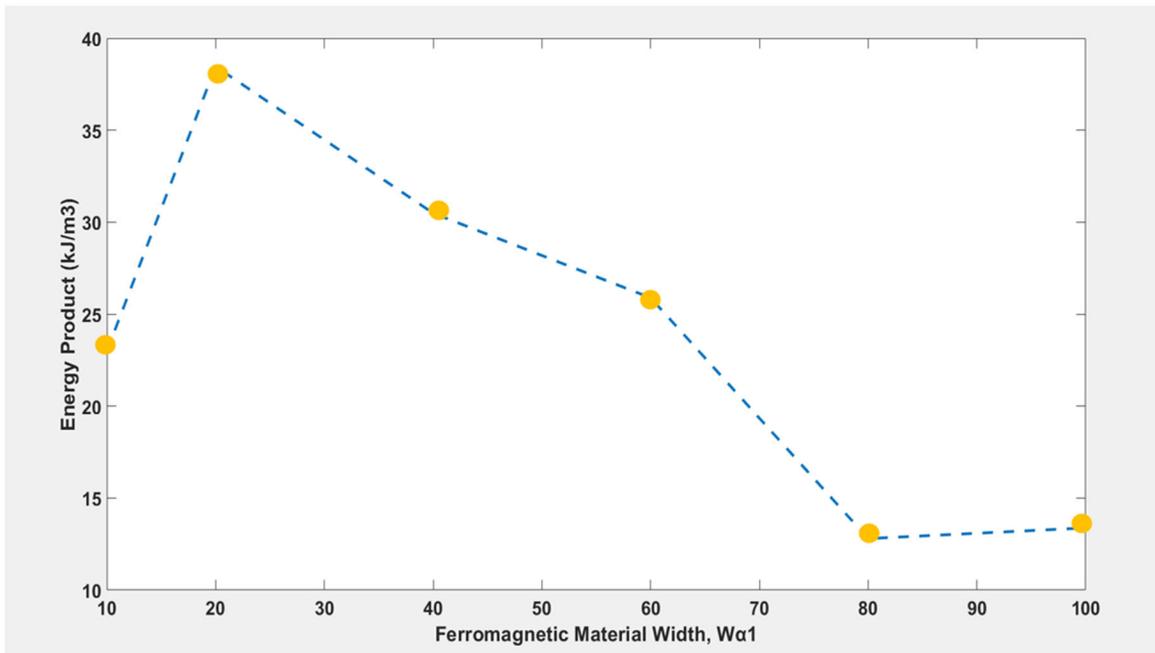

**Fig 4.** Variation in Energy Product $(BH)_{max}$ with increasing ferromagnetic material width ($W\alpha_1$). Here, the nonmagnetic material width is fixed at 40 nm.

Then, we studied magnetization reversal in Fig 5, where the geometry of 40 nm Wα1 (ferromagnetic material width) and 40 nm Wα2 (nonmagnetic material width) is taken. This geometry is chosen as a representative, as the magnetization reversal behavior is almost identical for all other geometry. Initially, all the spins are aligned in the positive y-direction at [0,1,0] (fig 5 b), and here, the applied external magnetic field, B_ext, is zero. As we start applying an external magnetic field, in the region of 0 to 500 mT and then 500 mT to 0 mT, it shows no significant change in spin direction; hence, spins keep their initial state, aligning collectively at [0,1,0] direction. As the applied external magnetic field turns negative, spins start to change their direction due to an applied field (significant change starts at around -150 mT (fig 5 c), and the spin reversal is completed at around -230 mT, eventually reaching the negative y direction [0,-1,0]. In summary, the top and bottom spins reversed quickly at around $B\_ext$ of -150 mT, the center spins responded slowly, and all the spins were completely reversed at around -230 mT (fig 5 e). Nevertheless, we do not observe stable multi-domain states during reversal, which explains why the $M_R$, $H_C$, and $(BH)_{max}$ are not significantly changed as a function of length upto 3 microns studies here. The external magnetic field is applied till -500 mT and the spins hold onto their negative y direction. Then we continue from -500mT to 500mT to complete the hysteresis loop cycle. As the positive external magnetic field is applied along the x-axis, the spins again show a spatially dependent behavior and start its reversal to the positive y-axis [0,1,0] at around 250mT. Overall, during the process, the center spins respond to higher values of $B\_ext$, while the top and bottom spins reverse at slightly lower $B\_ext$.

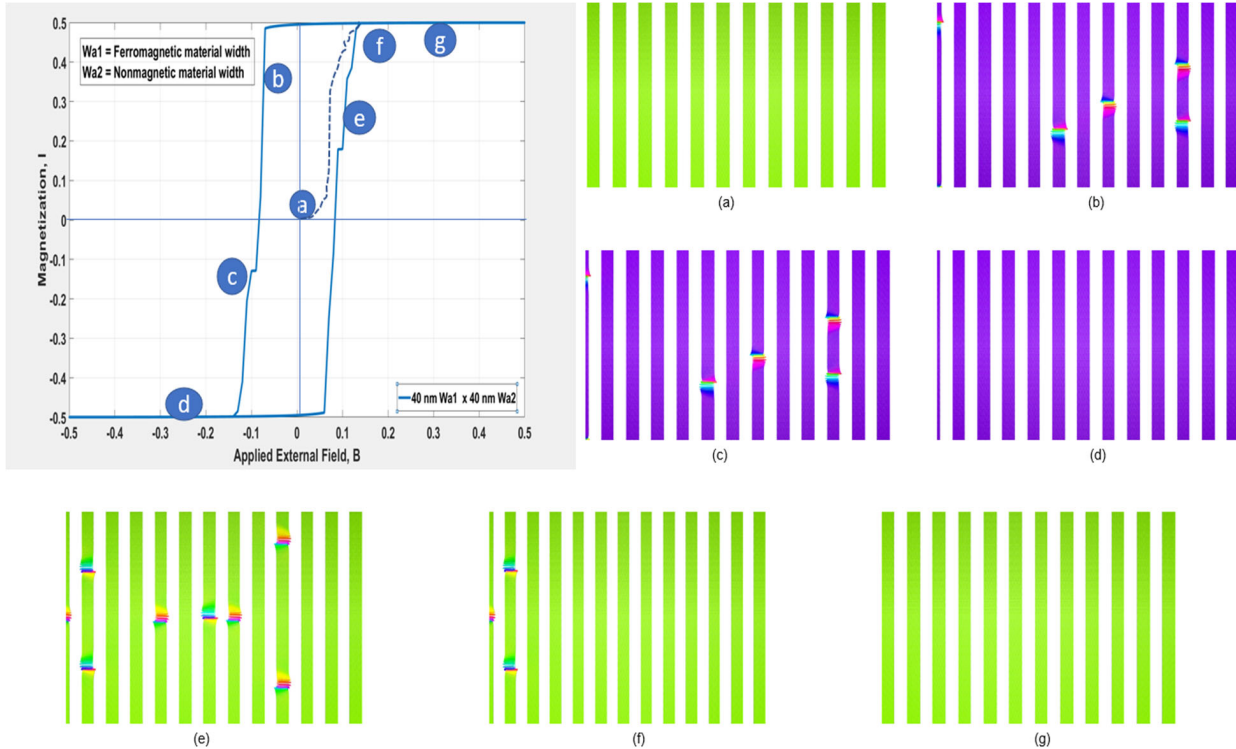

**Fig 5.** A demonstration of magnetization dynamics at geometry (40×40) nm, where both the ferromagnetic width and nonmagnetic material widths are 40 nm. Here, the green color represents that the spins are aligned in the positive y-axis, and the violet color represents that the spins are in the negative y direction. (b) initially, all the spins are aligned along (+) y-axis [0,1,0] (here, applied external magnetic field is zero), (c) spins start to rotate towards (-) y-axis at around $B\_ext$ of -150 mT, (d) at $B\_ext$ -230 mT spins completely aligned along (-)y-axis [0,-1,0], (f) At $B\_ext$ around 190 mT, the spins again start rotating towards the (+) y-axis, (g) at around $B\_ext$ 250 mT the spins complete their reversal and come back to the initial (b) position, fully aligning with (+) y-axis [0,1,0].

In Fig 6 (a), we consider a geometric configuration where both $W\alpha_1$, $W\alpha_2$, and thickness are constant at 40 nm and vary the ferromagnetic material strip length. Fig 6 (a) shows hysteresis loops with ferromagnetic material lengths of 3000, 1500, 1000 & 500 nm, respectively (the gap between two longitudinal ferromagnetic strips being 50 nm) where the same external magnetic field is applied (from -500 to +500 mT). The hysteresis loops show no significant changes, coercivity, remanence, and maximum energy product values are somewhat similar for all the cases (Fig 6 a). The lack of change in the maximum energy product is due to similar domain structures within each ferromagnetic segment.

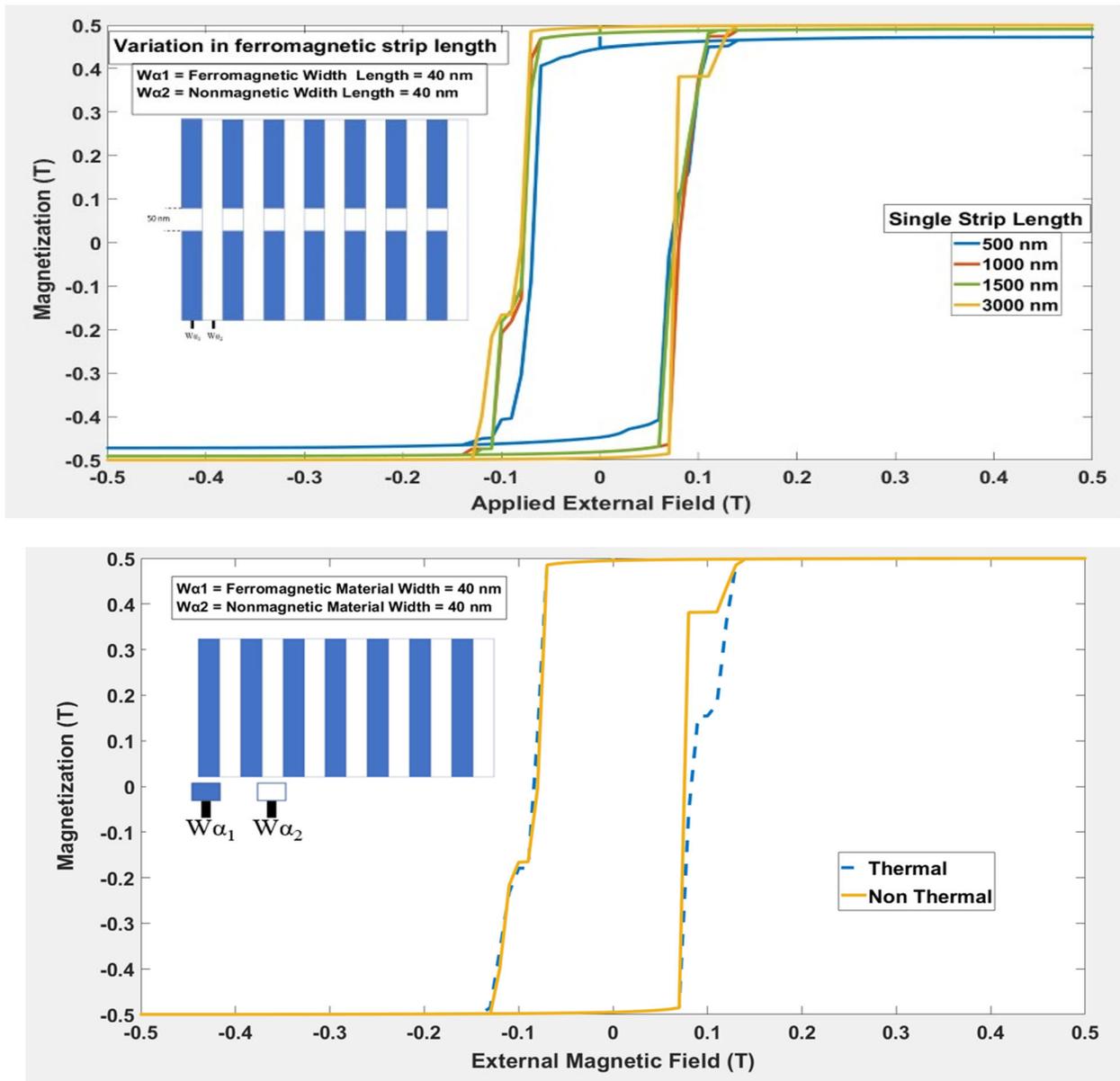

**Fig 6.** (a) For a specific size of $W\alpha_1$= 40 nm & $W\alpha_2$ = 40nm, hysteresis loops for different lengths of ferromagnetic material. In Fig 6(a), Blue, red, green, and yellow solid line loops represent ferromagnetic material single strip lengths of 500, 1000, 1500, and 3000 nm, respectively (the total length of the ferromagnetic region for all the cases is constant at 3000 nm), 6.(b) for size 3000 x 40 x40 nm3 thermal noise (T=300 k) is incorporated. here, $W\alpha_1$ (ferromagnetic material width) and $W\alpha_2$ (nonmagnetic material width) are 40 nm.

In Fig 6(b), we consider incorporating thermal noise to observe its effects on the hysteresis loop and energy products. Here, both the ferromagnetic and nonmagnetic material width is 40nm, and the length of the ferromagnetic material is 3000 nm. The blue dotted line hysteresis loop (fig 6(b)) is under thermal noise (T=300 K) and the other loop represented by the orange solid line is non-thermal. In the presence of thermal noise, the hysteresis curve transition is smoother along the edges as it's easier for domains to move out of pinning sites because the addition of random thermal energy provides additional activation energy needed to overcome the pinning barriers, which leads to increased domain wall mobility making the magnetic material more responsive to external magnetic fields. However, in terms of overall magnetic strength, the change due to the addition of thermal energy is negligible (less than 5% for ~300K, i.e. room temperature thermal noise).

**CONCLUSION**

In our investigation, we observed that as the width of ferromagnetic nanostructures increases, the influence of dipole interactions becomes evident (previous research articles also suggest the same thing), leading to a decrease in overall magnetic coercivity and, consequently, $(BH)_{max}$. However, a ferromagnetic inclusion width much lower than the non-magnetic gap leads to low $M_R$ and a smaller $(BH)_{max}$. Achieving an optimum packing factor becomes crucial for fully harnessing the maximum $(BH)_{max}$. Additionally, our study incorporates the presence of paramagnetic material (air) alongside permanent ferromagnets, laying down a valuable foundation for future research endeavors showing an ideal $(BH)_{max}$ of different geometries through which we can achieve 20% of the energy product of Rare Earth magnets. In the future, the integration of antiferromagnetic materials instead of airgaps, exchange coupled with ferromagnetic strips leading to negative exchange bias, along with inclusion of defects and roughness to increase coercivity and remnant magnetization holds promise for significantly impacting the magnetic strength and hence increasing the maximum energy product *$(BH)_{max}$* in rare-earth free magnets.

**ACKNOWLEDGEMENT**

The authors acknowledge US National Science Foundation Grant CMMI # 2310234.